\documentclass[12pt,preprint]{aastex}
\usepackage{graphicx}

\begin{document}

\title{The Revised Orbit of the $\delta$ Sco System}
\author{
C. Tycner,\altaffilmark{1}
A. Ames, \altaffilmark{1}
R.~T.~Zavala,\altaffilmark{2}
C.~A.~Hummel,\altaffilmark{3}
J.~A.~Benson,\altaffilmark{2}
D.~J.~Hutter\altaffilmark{2}
}

\altaffiltext{1}{Department of Physics, Central Michigan University,
  Mount Pleasant, MI 48859} 
\altaffiltext{2}{US Naval Observatory, Flagstaff Station, 10391
  W.~Naval Observatory Rd., Flagstaff, AZ 86001}
\altaffiltext{3}{European Southern Observatory,
  Karl-Schwarzschild-Str. 2, 85748 Garching
  bei M\"{u}nchen, Germany}

\slugcomment{\it Accepted to ApJ Letters}

\begin{abstract}
In anticipation of the possible collision between a circumstellar disk
and the secondary star in the highly eccentric binary system $\delta$
Scorpii, high angular resolution interferometric observations have
been acquired aimed at revising the binary parameters.  The Navy
Prototype Optical Interferometer (NPOI) was used to spatially resolve
the binary components in 2000 and over a period between 2005 and 2010.
The interferometric observations are used to obtain the angular
separations and orientations of the two stellar components at all
epochs for which data has been obtained, including 2005 and 2006, for
which based on previous studies there was some uncertainty as to if
the signature of binarity can be clearly detected.  The results of
this study represent the most complete and accurate coverage of the
binary orbit of this system to date and allow for the revised timing
of the upcoming periastron passage that will occur in 2011 to be
obtained.
\end{abstract}


\section{Introduction}

$\delta$ Scorpii (HD~143275, HR~5953, FK5~594) is a well known binary
system with a highly elliptical orbit and a period of almost 11
years. The primary star is classified as a Be star with a gaseous
circumstellar disk and the secondary is a B2-type
star~\citep{tango09}. This system has been a subject of a number of
studies that investigated a range of different characteristics ranging
from the disk structure around the primary~\citep{carciofi06,
  halonen08, millan10} to refining the binary
parameters~\citep{bedding93, hartkopf96, miro01, mason09, tango09}.
Because the primary did not possess strong H$\alpha$ emission during
the previous periastron passage in 2000, whereas as of late
$\delta$~Sco has shown strong H$\alpha$ emission, the high likelihood
that some type of interaction between the secondary and the
circumstellar disk of the primary will occur during the upcoming
periastron passage in 2011 makes this system particularly interesting.
Being such a high priority target for many observational campaigns
scheduled for 2011 creates the need for accurate binary parameters
such that the upcoming periastron passage can be predicted with high
accuracy.  This is especially important for the targeted observational
campaigns that request a very limited number of telescope nights near
the periastron passage.

The most recent binary parameters established for $\delta$~Sco were
presented by \citet{mason09} and \citet{tango09}.  \citet{mason09}
have revised the orbit using speckle interferometry complemented with
radial velocity data from \citet{miro01}.  Similarly, \citet{tango09}
combined the speckle data with additional optical interferometric data
from the Sydney University Stellar Interferometer (SUSI) obtained
between 1999 and 2007.  However, except for the observations obtained
on four nights in 1999, the interferometric observations obtained with
SUSI have not detected the binary signature.  \citet{tango09}
attributed this to a fully resolved circumstellar structure~(with an
estimated size of more than 4~mas) associated with the primary
component.  However, this is inconsistent with the results of
\citet{millan10} who based on 2007 observations estimated the size of
the circumstellar disk to be in the 1~mas range (although their
measurements have been obtained in the H and K band as opposed to the
optical).  In this study, we show how the interferometric observations
obtained with the Navy Prototype Optical Interferometer~(NPOI) at the
same epochs clearly resolve the binary signature. Nevertheless,
\citet{tango09} have obtained the best orbital elements published to
date with a newly revised period of 10.74~$\pm$~0.02~yr.

The high eccentricity of this system requires that the period should
be known with high accuracy if the periastron passage is to be
predicted with sufficiently high accuracy to be useful for targeted
observational campaigns.  This is because uncertainties of even a few
days in this 11-year period translate to very significant motions near
the periastron passage.  Therefore, to assist any upcoming
observational campaigns aimed at observing the $\delta$~Sco system
near its periastron passage, the main focus of this study is the
revision of the binary parameters.  We acquired interferometric
observations of the $\delta$ Sco binary system using the NPOI, which
are used to obtain high precision relative astrometric positions
between the stellar components.  The astrometric data is then combined
with the radial velocities from \cite{miro01} to produce the most
accurate orbital parameters for this system to date.

\section{Observations and Reductions}

The interferometric data used in this study have been acquired using
the NPOI on two nights in 2000 July (shortly before the last
periastron passage) and with the rest obtained over 94 nights between
2005 June and 2010 August.  There were about a dozen or so nights in
the latter period that have not been utilized in this study since
these nights produced small amounts of low quality data most commonly
due to poor atmospheric conditions.  Therefore, a total of 96 nights
have been used where the dates of all these nights are listed in
Table~\ref{tab:data}.

The NPOI is a six-element optical long-baseline interferometer that
has been described in detail by \citet{armstrong98}.  The
interferometric observations obtained during our $\delta$~Sco campaign
have utilized all six operational stations resulting in baseline
lengths from 18 to 80~m.  The raw interferometric data were obtained
and reduced using steps outlined in \citet{hummel03},
\citet{benson03}, and \citet{koubsky10}.  The calibrator star used was
$\zeta$~Oph (HR~6175, O9V) with an estimated diameter of 0.85~mas
based on the $R-I$ color~\citep{mozurkewich91}.

\begin{figure}
\plotone{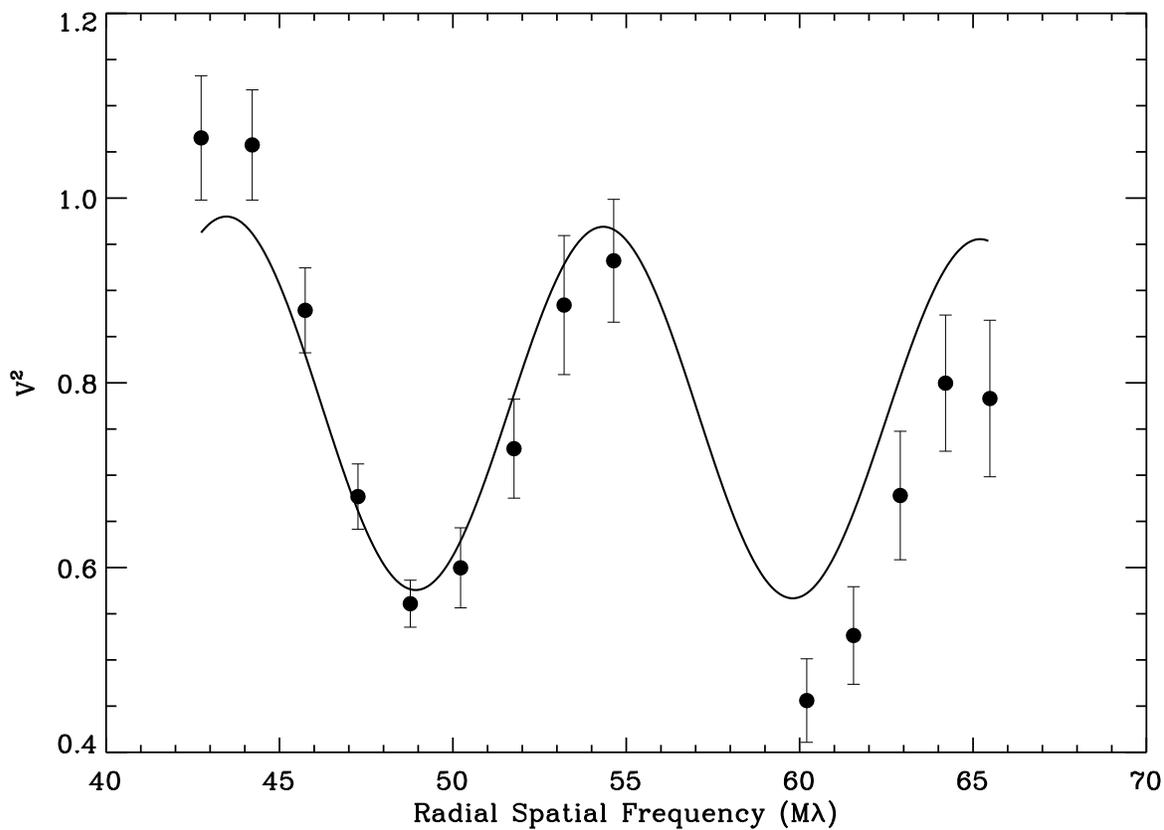}
\caption{Squared visibility data obtained for $\delta$~Sco on
  2006~June~5 across 14 spectral channels plotted as a function of
  radial spatial frequency.  Data from only one scan and one baseline
  are plotted for clarity to illustrate the interferometric binary
  signature in the form of a cosine~({\it solid line}) that is used to
  obtain the angular separation and orientation of the binary
  components.  }
\label{fig:binary}
\end{figure}

The interferometric binary signature for each individual night in our
data set was modeled in the same manner as described by equation~1 in
\citet{tango09}.  To estimate the squared visibility curve of the two
stellar components we assumed that the two angular diameters are
known.  For the primary component we adopted the measured uniform disk
diameter of 0.45$\pm$0.04~mas from \citet{hb74}.  For the secondary
component we adopted an angular diameter of 0.2~mas (nearly unresolved
for the employed baselines), which is expected
based on the average value of 5.7~$R_{\sun}$ from tabulations of
\citet{AQ} for a B2-type main sequence star and a dynamical parallax
of 7.03$\pm$0.15~mas \citep{tango09}.  Figure~\ref{fig:binary}
illustrates a typical interferometric binary signature across
different spectral channels of a single baseline obtained during the
2006 observing season.  Because the main purpose of this study is to
revise the binary parameters, we exclude from the analysis the
spectral channel that contains the H$\alpha$ emission from the
circumstellar disk around the primary.  We leave the analysis of the
H$\alpha$-emitting region in the $\delta$~Sco system for future
publication.

The resulting binary fits to each night produced the angular
separation ($\rho$) and the position angle (P.A., $\theta$) of the two
stellar components, as well as their magnitude difference.  The median magnitude
difference between the two components based on all the nights yielded
$\Delta m$ values of 1.87$\pm$0.17 and 2.24$\pm$0.26 for the 550 and
850~nm regions, respectively.  This is in an agreement with the values
of $\approx 1.9$ \citep{hb74} and 1.78$\pm$0.03 \citep{tango09}, based
on observations in the 440~nm region. The astrometric fits were repeated
adopting the median values of the magnitude differences, and the 
uncertainty ellipses were adopted as one-seventh the size of a 
Gaussian fitted to the center of the synthesized point-spread function
\citep{hummel03}. The astrometric results for all nights
are listed in Table~ \ref{tab:data}.

\section{Results}

\subsection{Revised Binary Parameters}

Following a similar procedure outlined by \citet{hummel03} we have
combined our astrometric measurements with the radial velocities
obtained by \citet{miro01} during the last periastron passage that
occurred in 2000.  Because \citet{miro01} did not publish uncertainty
estimates, we adopted 1.8 km~s$^{-1}$ as the measurement error of a
radial velocity based on their scatter in the final fit.  The
astrometric uncertainty ellipses did not have to be rescaled before
combination as their size corresponded to the average deviation from
the fitted positions.  For the systemic radial velocity we used the
value of $-7$ km s$^{-1}$ from \citep{evans67}. We also fixed the mass
of the secondary at $8 M_\odot$, the value found by \citet{tango09}.
Based on its color of $(B-V)=-0.20$ derived from the measured
magnitude differences, it is likely a dwarf of type B2 for which a
mass of $8 M_\odot$ is listed by \citet{AQ}.
This leaves the mass of the primary
and the seven orbital elements to be fit to the combined data
using standard nonlinear least-squares methods.
 
The combined
astrometric and radial velocity data resulted in the best-fit orbital
parameters listed in Table~\ref{tab:parameters}, where our values are
also compared to the results from \cite{miro01}, \cite{mason09}, and
\cite{tango09}.  The reduced $\chi^2$ statistic associated with the
best-fit solution was 1.2. The mass of the primary is derived as
$12.4\pm0.8 M_\odot$, and the dynamical parallax is $7.4\pm0.2$ mas.

\begin{figure}
\plotone{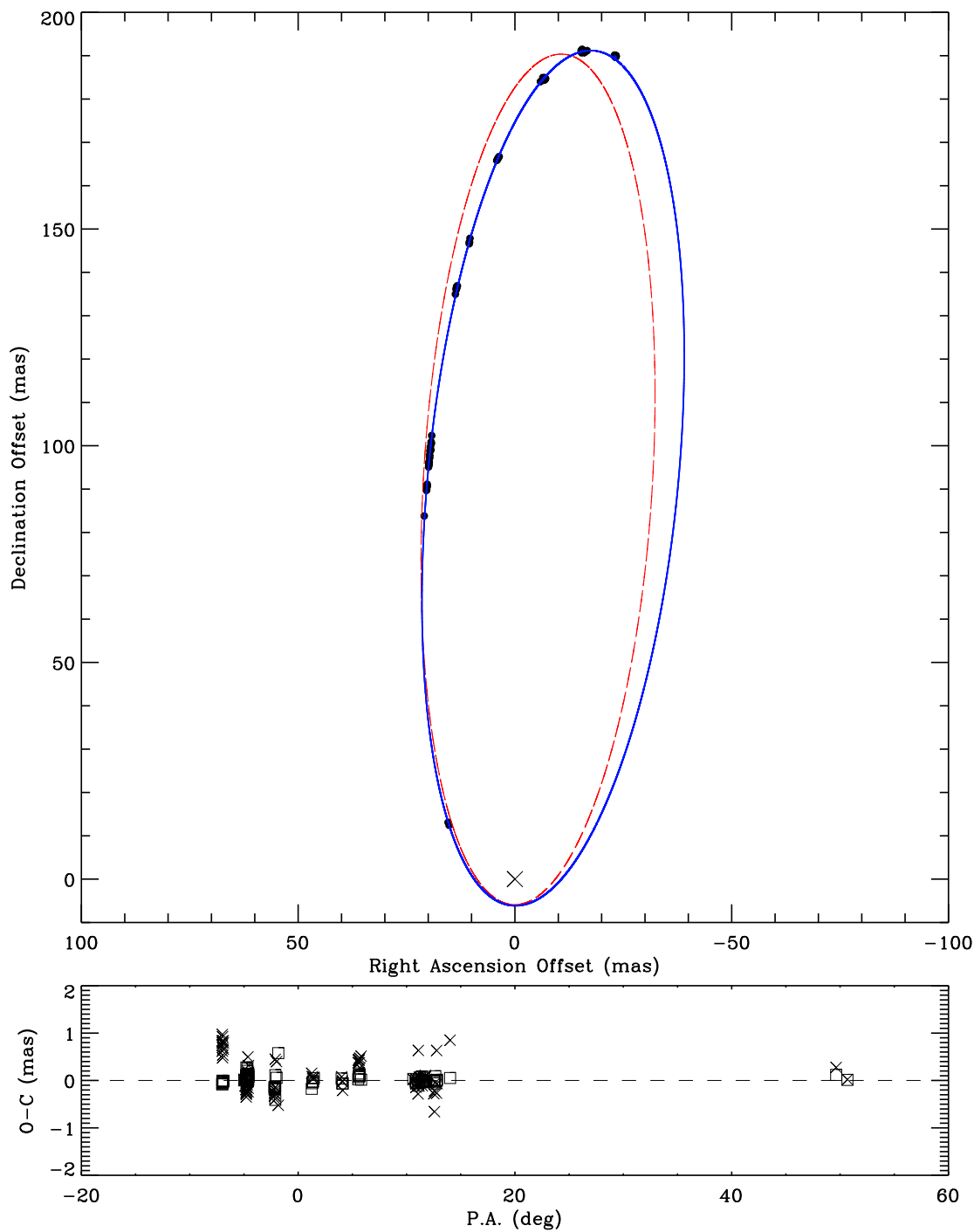}
\caption{Top panel: Binary orbits based on the parameters from
  Table~\ref{tab:parameters} based on the best-fit parameters from
  this study~({\it solid line}) and those based on Tango~et~al.~({\it
    dashed line}) plotted with the astrometric results ({\it filled
    circles}) of Table~\ref{tab:data}. The uncertainty ellipses of the
  astrometric data are generally smaller than the size of the plotted
  symbols and were omitted for clarity. The location of the primary is
  marked with an X. Bottom Panel: The east-west~({\it squares}) and
  north-south~({\it crosses}) components of the O--C vectors as a
  function of the P.A. of the secondary.}
\label{fig:orbit}
\end{figure}

Figure~\ref{fig:orbit} compares how well the revised model fits the
actual astrometric data and compares the revised orbit to the one
obtained by \citet{tango09}.  It is clearly evident from the figure
that the wide range of baselines of the NPOI produces very high
precision narrow-angle astrometric measurements.  The high precision
astrometry compounded with the extensive orbital coverage that
includes observations close to the previous periastron passage,
results in the best orbital parameters to date.

\subsection{Periastron Passage}

\begin{figure}
\plotone{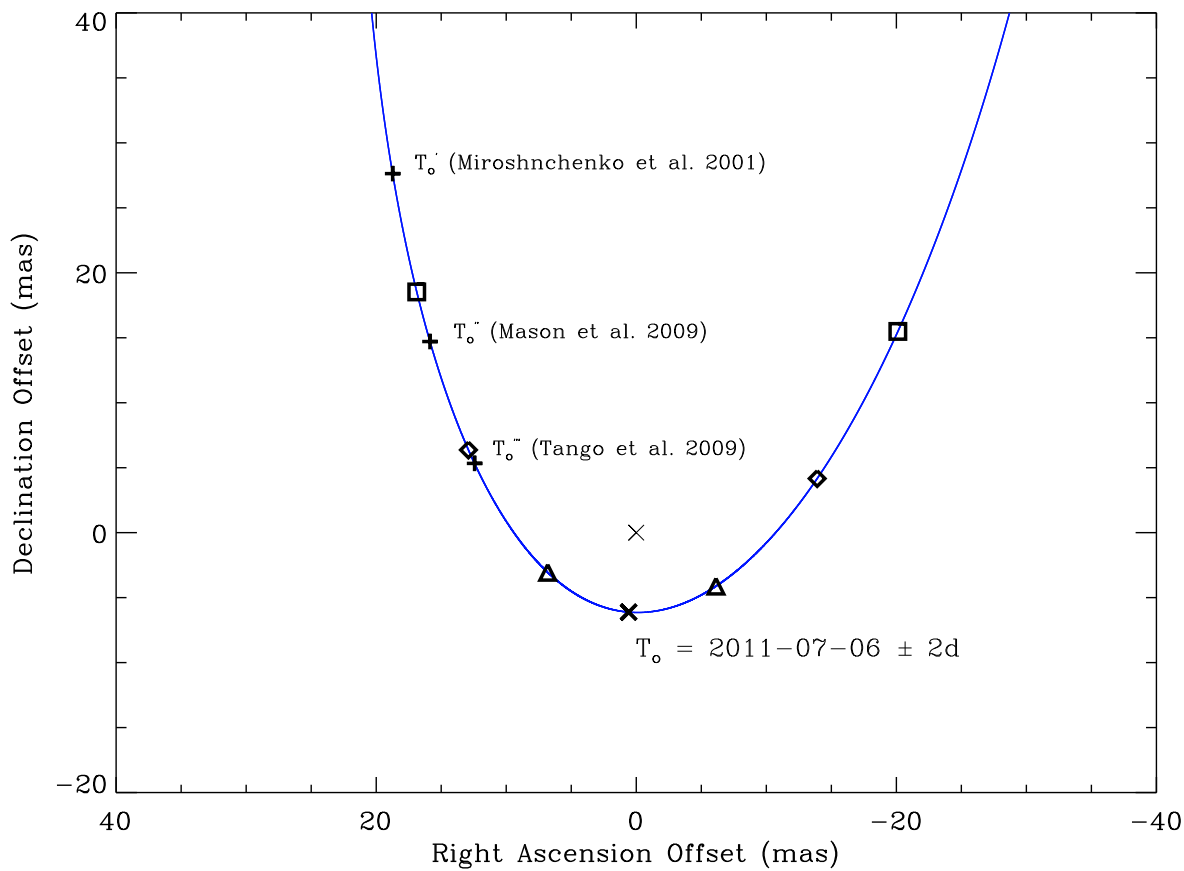}
\caption{Predicted positions of the secondary during the upcoming
  periastron passage in 2011 at 10 days~({\it triangles}), 30
  days~({\it diamonds}) and 60 days~({\it squares}) before and after
  the periastron passage~({\it cross}).  Locations of the secondary
  along the newly revised orbit based on periastron timings from
  previous studies are also shown~({\it pluses}).  All positions are
  measured with respect to the primary, which is located at the origin
  of the plot.}
\label{fig:periastron}
\end{figure}

Our best-fit values for $T_0$ and $P$ of JY2000.6927~(JD2,451,798) and
10.817~yr~(3950.9 $\pm$ 1.7~d), respectively, allow us to revise the
timing of the upcoming periastron passage to UT2011~Jul~6 $\pm$ 2~d.
This revised periastron timing differs by $\approx$~30 days from the
predictions based on the most accurate orbital parameters published to
date.  Because the secondary moves rapidly near periastron passage,
observations acquired even a few days before or after the closest
approach might miss the possible disk-secondary interaction
completely.  Figure~\ref{fig:periastron} illustrates where the
secondary is expected to be located 10, 30, and 60 days before and
after the periastron passage, along with the locations of the
secondary along our newly revised orbit based on the periastron
timings derived based on the $T_0$ and $P$ values published in the
literature.

Using the updated orbital parameters we also derive the minimum
expected separation at the closest approach~($\rho_{\rm min}$) to be
$6.14\pm 0.07$ mas (14 stellar radii).  The uncertainty in the
separation at the closest approach was obtained using a Monte Carlo
approach, which simulated 300,000 synthetic data sets all distributed
according to the uncertainties associated with all the orbital
parameters.  By obtaining a $\rho_{\rm min}$ value for each synthetic
data set allowed us to obtain a distribution of solutions where the
standard deviation of the distribution was used to obtain the
uncertainty on $\rho_{\rm min}$.

\section{Summary}

The binary orbit has been refined using astrometric data obtained from
NPOI and radial velocities from \cite{miro01}. The orbit that is
obtained gives a better fit to the data than previous findings. The
next periastron passage is expected to occur on UT~2011~Jul~6
$\pm$~2~d. Our results indicate that the periastron passage will
occur almost one month later compared to what one would expect based
on the period reported by \cite{tango09}. This result can have a
significant impact on observational campaigns that request relatively
small amount of telescope time (sometimes months in advance) and need
to time their request as closely to the periastron passage as
possible.

\small

\acknowledgements

The Navy Prototype Optical Interferometer is a joint project of the
Naval Research Laboratory and the US Naval Observatory, in cooperation
with Lowell Observatory, and is funded by the Office of Naval Research
and the Oceanographer of the Navy.  C.T. acknowledges, with thanks,
grant support from Research Corporation for Science Advancement.
C.T. and A.A. would like to thank the Physics Department at Central
Michigan University for support.

{\it Facilities:} \facility{NPOI}

\begin{table}[p]
\caption[]{\sc \small NPOI Astrometric Results}
\label{tab:data}
\begin{tabular}{lccccccc} \hline\hline
{UT Date}     & {JY}      & {\footnotesize $\rho$}  & {$\theta$} & \multicolumn{3}{c}{Error Ellipse}  \\ \cline{5-7}
{}            & {}        & {}        & {}         & $\sigma_{maj}$ & $\sigma_{min}$ & $\phi$  \\ 
{}            & {(yr)}    & {(mas)}   & {(deg)}    & {(mas)}        & {(mas)}        & {(deg)} \\ \tableline
2000 Jul 27 & 2000.5690 &  20.23 &  49.64 &  1.171 &  0.124 &  12.8\\
2000 Jul 28 & 2000.5718 &  19.62 &  50.68 &  0.963 &  0.113 &   8.5\\
2005 Jun 28 & 2005.4889 & 191.45 & 353.01 &  0.605 &  0.069 & 179.6\\
2005 Jun 29 & 2005.4917 & 191.30 & 352.99 &  0.575 &  0.061 & 175.3\\
2005 Jun 30 & 2005.4944 & 191.26 & 353.00 &  0.603 &  0.058 & 174.8\\
2005 Jul  1 & 2005.4971 & 191.15 & 353.01 &  0.556 &  0.065 & 178.4\\
2005 Jul  2 & 2005.4999 & 191.38 & 353.01 &  0.562 &  0.066 & 177.0\\
2005 Jul  3 & 2005.5026 & 191.48 & 353.03 &  0.621 &  0.062 & 175.4\\
2005 Jul  4 & 2005.5054 & 191.05 & 353.03 &  0.576 &  0.065 & 176.9\\
\hline
\end{tabular}\\[0.9ex]
\parbox{5.0in}{\footnotesize {\bf Notes.} Table~\ref{tab:data} is
  published in its entirety in the electronic edition of the
  Astrophysical Journal. Column 1: UT date of the observation; Column
  2: Julian year; Column 3-4: separation and P.A. (north through
  east); Column 5: semi-major axis of error ellipse; Column 6:
  semi-minor axis of error ellipse; Column 7: P.A. of error ellipse}
\end{table}

\begin{table}[p]
\caption{\sc \small The orbital elements for $\delta$ Sco}
\label{tab:parameters}
\begin{tabular}{lcccc}\hline\hline
Orbital        & Miroshnichenko        & Mason et al.       & Tango et al.             & This Work \\
Parameters     & et al. (2001)         & (2009)             & (2009)                   &\\
\tableline
$a$ (mas)      & $107^{\rm a}$         & $104 \pm 6$        & $98.3 \pm 1.2$           & $99.1 \pm 0.1$ \\
$i$ (deg)      & $38 \pm 5$            & $39 \pm 8$         & $38 \pm 6$               & $32.9 \pm 0.2$ \\
$\Omega$ (deg) & $175^{\rm a}$         & $153 \pm 9$        & $175.2 \pm 0.6$          & $172.8 \pm 0.9$ \\
$e$            & $0.94 \pm 0.01$       & $0.94^{\rm b}$     & $0.9401 \pm 0.0002$      & $0.9380 \pm 0.0007$\\
$\omega$ (deg) & $-1 \pm 5$            & $29 \pm 12$        & $1.9 \pm 0.1$            & $2.1 \pm 1.1$ \\
$T_0$ (JY)     & $2000.693 \pm 0.008$  & $2000.693^{\rm b}$ & $2000.69389 \pm 0.00007$ & $2000.6927 \pm 0.0014$ \\
$P$ (year)     & $10.58^{\rm a}$       & $10.68 \pm 0.05$   & $10.74 \pm 0.02$         & $10.817 \pm 0.005$ \\
\hline
\end{tabular}\\[0.9ex]
\parbox{6.0in}{\footnotesize {\bf Notes.}\\
 $^{\rm a}${Parameter adopted from \cite{hartkopf96} solution.}\\
 $^{\rm b}${Parameter adopted from \cite{miro01} solution.}\\
}
\end{table}


\end{document}